\documentclass[conference]{IEEEtran}
\usepackage[fleqn]{amsmath}
\usepackage{amsthm}
\usepackage{amssymb}
\usepackage{msc}
\usepackage{xspace}
\usepackage{paralist}
\usepackage{balance}
\usepackage{booktabs}
\usepackage{tabularx}
\usepackage{longtable}
\usepackage{listings}
\usepackage{cite}               %
\usepackage{breakurl}           %
\usepackage{url}                %
\usepackage{xcolor}             %
\usepackage[hidelinks]{hyperref}         %

\usepackage[utf8]{inputenc}

\newtheorem{definition}{Definition}

\ifdefined\showcomments
\usepackage{todonotes}
\usepackage{letltxmacro}
\LetLtxMacro{\todonote}{\todo}
\renewcommand{\todo}[2][]
{\todonote[inline, caption={#2}, size=\footnotesize, #1]
{\renewcommand{\baselinestretch}{0.5}\selectfont#2\par}}
\else
\usepackage[disable]{todonotes}
\fi

\newcommand{\sys}{\texttt{VICEROY}\xspace}

\newcommand{\vcr}{VCR\xspace}
\newcommand{\wrapper}{wrapper\xspace}
\newcommand{\Wrapper}{Wrapper\xspace}

\newcommand{\taggedpara}[1]{\noindent\textbf{#1.}}

\definecolor{mygreen}{rgb}{0,0.6,0}

\author{
\IEEEauthorblockN{Scott Jordan}
\IEEEauthorblockA{UC Irvine\\
sjordan@uci.edu}
\and
\IEEEauthorblockN{Yoshimichi Nakatsuka\IEEEauthorrefmark{1}}
\IEEEauthorblockA{UC Irvine\\
nakatsuy@uci.edu}
\and
\IEEEauthorblockN{Ercan Ozturk\IEEEauthorrefmark{1}}
\IEEEauthorblockA{UC Irvine\\
ercano@uci.edu}
\and
\IEEEauthorblockN{Andrew Paverd}
\IEEEauthorblockA{Microsoft Research\\
andrew.paverd@microsoft.com}
\and
\IEEEauthorblockN{Gene Tsudik}
\IEEEauthorblockA{UC Irvine\\
gene.tsudik@uci.edu}
}

\begin{document}

\title{\sys: GDPR-/CCPA-compliant Enforcement of \\Verifiable Accountless Consumer Requests}

\IEEEoverridecommandlockouts
\makeatletter\def\@IEEEpubidpullup{6.5\baselineskip}\makeatother
\IEEEpubid{\parbox{\columnwidth}{
    Network and Distributed System Security (NDSS) Symposium 2023\\
    27 February - 3 March 2023, San Diego, CA, USA\\
    ISBN 1-891562-83-5\\
    \url{https://dx.doi.org/10.14722/ndss.2023.23074}\\
    \url{www.ndss-symposium.org}
}
\hspace{\columnsep}\makebox[\columnwidth]{}}

\maketitle
\begingroup\renewcommand\thefootnote{}
\footnotetext{\IEEEauthorrefmark{1} Corresponding authors. Authors ordered alphabetically.}
\endgroup

\begin{abstract}
Recent data protection regulations (notably, GDPR and CCPA) grant consumers various rights, 
including the right to \emph{access}, \emph{modify} or \emph{delete} any personal information collected about them (and retained) by a service provider.
To exercise these rights, one must submit a \emph{verifiable consumer request} proving that the collected data indeed pertains to them.
This action is straightforward for consumers with active accounts with a service provider at the time of data collection, since they can use standard (e.g., password-based) means of authentication to validate their requests.
However, a major conundrum arises from the need to support consumers \emph{without accounts} to exercise their rights.
To this end, some service providers began requiring such {\em accountless} consumers to reveal and prove their identities (e.g., using government-issued documents, utility bills, or credit card numbers) as part of issuing a verifiable consumer request.
While understandable and reasonable as a short-term fix, this approach is cumbersome and expensive for service providers as well as privacy-invasive for consumers. 

Consequently, there is a strong need to provide better means of authenticating requests from accountless consumers.
To achieve this, we propose \sys, a privacy-preserving and scalable framework for producing \emph{proofs of data ownership}, 
which form a basis for verifiable consumer requests.
Building upon existing web techniques and features, \sys allows accountless consumers to interact with service providers, and later prove that they are the same person in a privacy-preserving manner, while requiring minimal changes for both parties.
We design and implement \sys with emphasis on security/privacy, deployability, and usability.
We also assess its practicality via extensive experiments. 

\end{abstract}
\section{Introduction}\label{sec:introduction}
Several new data protection regulations have been enacted in recent years, notably the European Union 
General Data Protection Regulation (GDPR)~\cite{GDPR} and the California Consumer Privacy Act (CCPA)~\cite{CCPA}. 
These regulations grant consumers various new legal rights. 
For example, consumers gain the right to \emph{access} personal data collected about them and held by service providers 
(GDPR Art.\ 15, CCPA 1798.100), \emph{request correction} (GDPR Art.\ 16, CCPA 1798.106) or 
\emph{request deletion} of their personal data (GDPR Art.\ 17, CCPA 1798.105).

Importantly, these regulations expand the definition of \emph{personal data} beyond that associated with a person's 
real name. For example, GDPR Rec.\ 30 states that natural persons ``\emph{may be associated with online identifiers 
provided by their devices, applications, tools and protocols, such as internet protocol addresses, cookie identifiers or 
other identifiers}''. This means that any website\footnote{As shorthand, we use the term 
\emph{website} to represent the entity operating a website, which (we assume) falls into the category of entities that the 
GDPR and CCPA call \emph{controller} and \emph{business}, respectively.} collecting information about consumers\footnote{We use \emph{consumer} or \emph{client} to refer to:
(1) the GDPR term \emph{data subject}, (2) the CCPA term \emph{consumer}, and (3) the equivalent terms in 
other regulations.} 
based on identifiers, such as IP addresses or cookies, may be collecting personal information, and thus have to comply with 
these new regulations.
The website must therefore provide a means by which consumers can access, request correction of, or 
request deletion of their personal information.

When dealing with a consumer request, the website must verify that the requestor is indeed the consumer 
to whom the personal information pertains. This is critical to prevent erroneous disclosure (which would be a serious violation of 
any data protection regulation), unauthorized modification, or deletion of personal information. %
This is called a ``\emph{verifiable consumer request}'' (\vcr) (CCPA 1798.140(y)).\footnote{These requests are 
sometimes also referred to as Subject Rights Request (SRR) or Subject Access Request (SAR).}

For consumers who have pre-existing accounts on a given website, submitting a \vcr is relatively 
straight-forward. To wit, CCPA 1798.185 requires: ``\emph{treating a request submitted through a password-protected 
account maintained by the consumer ... as a verifiable consumer request}''. 
However, there remains a major challenge of how to support a \vcr from casual or {\bf accountless} consumers, 
as required by CCPA 1798.185, whilst protecting such consumers' privacy.

The only mechanisms that are currently suitable for accountless consumers are those that require: a device 
cookie, a government-issued ID, a signed (and possibly witnessed) statement, a utility bill, a credit card number, 
or taking part in a phone interview~\cite{pavur2019gdparrrrr}. However, these mechanisms are cumbersome 
(and some are time-consuming) for consumers as well as insecure, as demonstrated by prior 
work~\cite{pavur2019gdparrrrr,cagnazzo2019gdpirated,di2019personal,boniface2019security}.
They typically require manual processing, which is both error-prone and costly.
Moreover, such methods (apart from device cookies) are privacy-invasive for the consumer and 
open the door for further consumer data exposure. For example, a government-issued ID or utility bill 
reveals even more private information to the website.

In light of these issues, the most appealing choice appears to be the use of device cookies.
Cookies are already used pervasively by websites to link multiple sets of activities (sessions) to the same consumer.
At first glance, asking for device cookies as part of a \vcr appears to meet the GDPR and CCPA requirements: only the authorized consumer should possess the correct cookie (\emph{unforgeability}), and providing a cookie does not reveal additional information (\emph{privacy}). 
However, this essentially means treating device cookies as \emph{authentication tokens}, which has at least three disadvantages:

First, in the general case, there is no requirement for a cookie's value to be unguessable.
A recent large-scale study~\cite{Gonzalez2017} found cookie values containing URLs, email addresses, timestamps, and even JSON objects.
Although \emph{authentication cookies} are designed to be unguessable, these would typically only be used once the user has logged into an account.

Second, cookies are used (i.e., sent over the network) whenever the consumer interacts with the website.
Although secure communication channels (e.g., TLS) can protect cookies in transit, the MITRE ATT\&CK framework lists several recent examples of techniques for stealing web cookies~\cite{mitre-cookies}.

Third, consumers must protect the cookies stored on their devices, especially considering e.g., client-side spyware, even after the cookies expire.
Secure storage may become more challenging as the number of stored cookies increases, since consumers should not delete cookies for which they might subsequently issue a \vcr. 
If an adversary can guess or obtain the cookie through any of the above vectors, they would be able to request, modify, or delete all the consumer's data.

Motivated by aforementioned issues, we construct \sys{}, a first-of-its-kind framework that allows {\em accountless} consumers to request their data in a private manner, while allowing website operators to efficiently and securely verify such requests.
\sys introduces a one-to-one mapping between Web sessions and consumer-generated public keys.
At session initiation, the consumer generates a public key (a \vcr public key) and supplies it to the server.
At a later time, the consumer digitally signs its request using the private key corresponding to the \vcr public key for the session.

To ensure consumer privacy, our key derivation mechanism uses unlinkable public keys derived from a single master public key.
This also allows \sys to only require consumers to securely store a single private key, regardless of the number of sessions they have generated.
Moreover, since this private key is only needed when generating \vcr{s}, it can be protected using well-known secure key storage mechanisms (e.g., hardware security device).

\sys is composed of well-known cryptographic primitives.
However, to meet the necessary requirements of security, scalability, and privacy, this must done through the careful selection of such primitives.
Furthermore, \sys's design prioritizes deployablity, requiring only minimal changes to existing websites and no changes to existing cookie usage.

The contributions of this work are:
\begin{compactitem}
	\item Design of \sys -- a secure, scalable, and privacy-preserving \vcr mechanism for accountless consumers.
	\item Careful selection of cryptographic protocols to balance the three requirements of \sys.
	\item Proof-of-concept implementation of \sys for web browsers, including a \sys-compatible hardware security device.
	\item Thorough security, performance, and deployability evaluation of \sys.
\end{compactitem}

\taggedpara{Organization} 
Section~\ref{sec:background} presents background on GDPR/CCPA and verifiable consumer requests (\vcr{s}).
Next, Section~\ref{sec:system_and_threat_models} presents our threat model and defines requirements for \sys.
Sections~\ref{sec:design_and_challenges} and~\ref{sec:implementation} then describe the design and 
proof-of-concept implementation of \sys{}. Section~\ref{sec:evaluation} presents our evaluation methodology and results.
Further aspects of \sys{} are discussed in Section~\ref{sec:discussion} and related work is overviewed in 
Section~\ref{sec:related_work}. Section~\ref{sec:conclusion_and_future_work} concludes the paper.

\taggedpara{Code Availability} Source code for all \sys components and the Tamarin model is available at~\cite{code}.
\section{GDPR/CCPA Background} 
\label{sec:background}
This section overviews Personally Identifiable Information 
and consumer rights under the GDPR and the CCPA. Given familiarity with GDPR and CCPA, 
it can be skipped without any loss of continuity.  

\subsection{Personally Identifiable Information (PII)}
Both GDPR and CCPA pertain to the combination of \emph{personal and personally identifiable} information, often referred to as: {\em Personally Identifiable Information}\footnote{The GDPR uses the term \emph{personal data} and the CCPA uses \emph{personal information.}} or PII.

Information is \emph{personal} if it relates to a person, e.g., contact information, geolocation, applications and devices used, how an application is used, interests, websites visited, consumer-generated content, identities of people with whom a consumer communicated, content of communication, audio, video, and sensor data~\cite{JordanSurvey}.

Information is \emph{personally identifiable} if the person to which it pertains is either identified or identifiable.
If information is paired with a name, telephone number, email address, government-issued identifier, or postal address, then it is considered to be personally identifiable~\cite{JordanCCPA}.
If information is paired with an IP address, a device identifier (e.g., an IMEI), or an advertising identifier, it is \emph{likely} to be considered as personally identifiable~\cite{JordanCCPA}.
Information paired with an identifier created by a business (e.g., a cookie) is personally identifiable if it can be combined with other information to allow the consumer to whom it relates to be identified~\cite{JordanCCPA}.

\subsection{Rights of Access and Erasure}
Both GDPR and CCPA require a business that collects PII to disclose, typically in its privacy policy, the 
categories of PII collected, the purposes for collecting it, and the categories of entities with which that PII is 
shared~\cite{JordanCCPA}. Both regulations give consumers the right:
\begin{compactitem}
\item~To learn about, and control, information relating to them 
that a business has collected. Specifically, consumers have the right to request access to the 
\emph{specific  pieces of PII} that the business has collected (GDPR Art. 15; CCPA Sec. 1798.110(a)(5)).
\item~To request that their incorrect PII be corrected (GDPR Art. 16; 
CCPA Sec. 1798.106). 
\item~To request that a business delete their PII (GDPR Art. 17; CCPA Sec. 1798.105).
\end{compactitem}

\subsection{Verifiable Consumer Requests (\vcr{s})}
The consumer rights to access, and request correction or deletion of, their PII are contingent upon verification that 
the consumer is indeed the person to whom that PII relates. However, GDPR and CCPA differ in requirements 
of methods of verification. Both regulations require a business to use reasonable measures to verify the consumer's identity 
(GDPR Rec. 64; CCPA Sec. 1798.140(ak)). If a consumer has a password-protected account with a business, both require 
a business to treat requests submitted via that account as verified (GDPR Rec. 57; CCPA Sec. 1798.185(a)(7)).

However, both regulations also recognize that PII is often collected about casual consumers, who do not have 
password-protected accounts. 
In this case, they envision a consumer request being verified by associating additional consumer-supplied 
information with PII that the business previously collected about that consumer (CCPA Sec. 1798.130(a)(3)(B)(i)). 
The CCPA further specifies that any information provided by the consumer in the request can be used solely for the purposes of 
verification (CCPA Sec. 1798.130(a)(7)). However, if a business has not linked PII to a consumer or a household, 
and cannot link it without the acquisition of additional information, then neither the GDPR nor the CCPA require a business to 
acquire additional information to verify a consumer request (GDPR Rec. 57; CCPA Sec. 1798.145(j)(3)). 
Thus, some requests may be unverifiable.

The CCPA~\cite{CCPA_regs} recognizes that consumer verification is not absolutely certain. It establishes two thresholds 
of certainty. The lower threshold, called \emph{reasonable degree of certainty}, may be satisfied by matching at least two 
pieces of information provided by the consumer (CCPA Regs. \S{}999.325(b)).
The higher threshold, called \emph{reasonably high degree of certainty}, may be satisfied by matching at least three pieces of information provided by the consumer, and obtaining a signed declaration from the consumer (CCPA Regs. \S{}999.325(c)).
However, other means of verification may also satisfy these thresholds.
Verification of the consumer identity must always, at a minimum, meet the {\em reasonable degree of certainty} threshold.
Furthermore, requests to learn specific pieces of PII must meet the {\em reasonably high degree of certainty threshold}.
Finally, a consumer may choose to use a third-party verification service (CCPA Regs. \S{}999.326).

The design of a verification method should balance the administrative burden on the consumer (CCPA Sec. 1798.185(a)(7)) 
with the likelihood of unauthorized access and the risk of harm (CCPA Regs. \S{}999.323(b)(3)).

\section{Threat Model and Requirements} \label{sec:system_and_threat_models}
Our system model assumes a typical Web environment with two types of principals: \emph{(1)} clients and \emph{(2)} servers.
Clients are {\em consumers} who access Internet services offered by servers.
Servers collect and store data during the interactions with the consumers by associating such data with identifiers issued to the consumer.
Each client can own multiple devices and at least one of the client's devices can be trusted to store a secret, e.g., a private key.
This trusted device could be a smartphone, a dedicated key storage device, or a secure hardware wallet.
All access to the secret is controlled by the client.
Physical and side-channel attacks against the trusted client device are beyond the scope of this paper.

We assume secure communication channels between clients and servers, which can be realized using standard means, e.g., HTTPS.
Use of secure channels to deliver web content has become a de-facto standard, as shown by Felt et al.~\cite{felt2017measuring}, 
which reports that up to 87\% of all webpages were served via HTTPS in 2017. This number is expected to increase, as shown by 
Google's 2022 Transparency Report~\cite{google2022https}, which claims that 80--98\% of top-100 websites use HTTPS.
Moreover, standards such as DTLS~\cite{rfc9147} and QUIC~\cite{rfc9000} allow devices that cannot use TCP to 
establish similarly secure channels.

We consider three types of adversarial behavior:
\begin{compactitem}
	\item \textbf{Malicious clients:} Attempt to impersonate other clients in order to perform operations on data that is not theirs.
	\item \textbf{Client-side malware:} Attempts to perform unauthorized operations on client data without client's knowledge. 
	We assume that the client's trusted device is free of malware, while all other client devices can be potentially infected.
	\item \textbf{Honest-but-curious servers:} Attempt to identify clients who submit requests, or
	to link multiple requests to the same client.
	Multiple servers might collude to link client requests and/or to learn client identities. 
\end{compactitem}
As usual, we assume all relevant cryptographic primitives are implemented and used correctly and cannot 
be attacked via side-channels or any other weaknesses. Similarly, we assume digital signatures  
can only be generated by the true owner of the private key.

Based on the above system model, we define the following requirements for \sys:
\begin{compactitem}
	\item \textbf{Unforgeability:} Only the client who originally interacted with the server can create a valid \vcr.
	\item \textbf{Replay resistance:} A server will only accept a valid \vcr at most once.
	\item \textbf{Consumer Privacy:} An honest-but-curious server (or a set thereof) should be unable to link a \vcr to a specific client, or to link multiple \vcr{s} to the same client. %
\end{compactitem}

\section{\sys Design \& Challenges}
\label{sec:design_and_challenges}
This section discusses \sys's goals, design features, and challenges encountered.
Note that we use the terms \emph{client} and \emph{consumer} interchangeably.

\subsection{Design Motivation}
\label{subsec:design_motivation}
One straight-forward way to support \vcr{s} from accountless consumers is to require them to provide the same cookie(s) they were issued when originally visiting the server website.
(Indeed this is one of the mechanisms that~\cite{pavur2019gdparrrrr} encountered in their survey of how businesses respond to access requests.)
The rationale is that only the consumer from whom the data was collected should have access to the cookie, which ties all consumer's activity that constitutes one session. 
This method has several advantages:
First, it is \emph{privacy-preserving} in that, when making a request, the consumer does not reveal any further personal information that the server didn't already have. 
Furthermore, if the consumer submits multiple \vcr{s} based on different cookies, the server cannot link them.\footnote{Potential ``fingerprinting'' of the consumer's browser or network interface notwithstanding.}
Second, this mechanism is easily deployable, since cookies are supported by virtually every device that uses the Web.

However, per Section~\ref{sec:introduction}, there are also several significant disadvantages:
First, this method essentially makes cookies into \emph{symmetric} authentication tokens: anyone in possession of the cookie can create \vcr{s}.
This is problematic because cookies, in general, are not required to be unguessable and may contain predictable information, such as URLs or email addresses~\cite{Gonzalez2017}.
A subset of cookies, namely \emph{authentication cookies}, are designed to be unguessable, but these would typically only be used once the consumer has logged into an account (i.e., no longer an accountless consumer).
Second, cookies are used in all interactions with the website, and several techniques for stealing cookies have been demonstrated~\cite{mitre-cookies}.
Third, since consumers visit many different websites, they would have to {\em securely} and {\em reliably} store a potentially large number of cookies.
This differs from the usual client-side cookie management, since cookies would be additionally valuable as a means to issue \vcr{s}.
Also, if cookies are lost (e.g., due to disk failure), the consumer would be unable to exercise their GDPR/CCPA rights. 
This underscores the importance of cookie storage reliability.
Furthermore, if the server for any reason also stores copies of cookies, same security requirements apply.
If these cookies are leaked as a result of a data breach, the server would have to invalidate them in bulk, thus preventing legitimate consumers from submitting \vcr{s}, or risk attackers requesting consumers' personal data.

\subsection{Conceptual Design}
\label{subsec:conceptual_design}
Motivated by aforementioned challenges, we construct \sys{} to avoid the drawbacks discussed. 
We now describe key features of \sys.

\taggedpara{Asymmetric tokens}
When interacting with a server, a client provides the server with the public part of an asymmetric key-pair called the \emph{\vcr{} public key}.
Upon receiving a \vcr{} public key, the server associates this key with a particular \emph{session}.
Generally, a session is any linkable set of interactions between a client and a server.
For example, in the Web context, a session most likely corresponds to an HTTP(S) session, which is managed using cookies.
To protect the client's privacy, a new \vcr{} public key, which is unlinkable to any previous keys, can be used for each session.
Finally, to submit a \vcr{} for a particular session, the client creates a request and signs it using the corresponding \vcr{} private key for that session.

This approach addresses the drawbacks of using only cookies to authenticate the request, since the client's signature is assumed to be unforgeable and the client's \vcr{} private keys are never sent over the network.
It does not matter if the adversary learns the \vcr{} public keys.
The use of digital signatures also allows additional information/parameters to be cryptographically bound to the request (e.g., a request to correct personal information could, in some cases, already include the corrected information).

\taggedpara{Cookie \wrapper{s}}
At first glance, mapping data collected during a session to the \vcr{} public key seems to be an efficient way of storing such data at server side, especially when the consumer submits a \vcr{}.
However, from a deployability perspective, it would be infeasible to replace existing Web cookies with public keys because this would require non-trivial modifications to the way servers use cookies.
For example, cookie values containing URLs, email addresses, timestamps, and even JSON objects have been observed in practice~\cite{Gonzalez2017}.

To avoid changing the existing and ubiquitous cookie mechanism, \sys{} introduces the concept of a \emph{cookie \wrapper{}} -- a cryptographic binding between an existing server-generated session identifier (e.g., cookie) and a client-generated \vcr{} public key.
The server generates a cookie \wrapper by signing the hash of the server-generated cookie and the \vcr{} public key using the server's long-term \wrapper signing key.
This allows the client to verify whether the \wrapper was created correctly.
This \wrapper is created contemporaneously with the cookie, and at most one \wrapper{} is created per cookie.
The \wrapper{s} are then sent to and stored by the client alongside the cookie and \vcr{} public key.
The use of cookie \wrapper{s} significantly improves deployability by allowing servers to add support for \sys{} without modifying current cookie management.

\taggedpara{Submitting VCRs}
When the client issues a \vcr{} signed with the relevant \vcr{} private key (as described above), the client also sends the corresponding cookie \wrapper{} to the server, along with the request.
The server first verifies that the \wrapper{} is valid, by verifying its own signature on the \wrapper{}.
If valid, the server then uses the \vcr{} public key specified in the \wrapper{} to verify the client's signature over the request.
If this in turn is valid, the server is assured that this request was generated by the same client who received the original cookie (i.e., the legitimate consumer).

\subsection{Design Challenges and Solutions} \label{subsec:design_challenges}
The conceptual design described above presents several design challenges.
This section outlines the main challenges and presents the key insights used to realize \sys{}.

\subsubsection{Avoiding Key Explosion}
\label{subsubsec:key_generation}
For privacy reasons, the client cannot use a static public key. 
A static public key would allow a server (or a set thereof) to link together multiple sessions by the same client.
This linkage could take place at session initiation time, i.e., when the client requests a \wrapper{}, or when the client issues a \vcr{}.
Also, if a client's static public key is leaked, it becomes possible to track that client's sessions globally.
However, requiring each client to have a distinct public/private key-pair per session can cause a ``key explosion'' in which the client has to manage the large quantity of public keys and more importantly, securely store many private keys.

To avoid this issue, we use the concept of \emph{derivable asymmetric keys}, specifically, the key derivation scheme used in Bitcoin Improvement Proposal (BIP) 32~\cite{bip32}.
This type of key derivation scheme allows a chain of \emph{child public keys} to be derived from a single \emph{parent public key}.
Importantly, the derivation of public keys does not require access to the corresponding parent private key.
Furthermore, the corresponding child private keys can only be derived from the parent/master private key.
We denote the \emph{derivation path} of a key as \texttt{a/b/c/...}, where \texttt{a/b} is the $b^{th}$ child key of \texttt{a}, and \texttt{a/b/c} is the $c^{th}$ child key of \texttt{a/b}.
This approach minimizes public key storage requirements of \sys{} -- only the parent public key must be stored, whilst all other public keys can be derived.
When a new session is initiated, the parent public key is used to generate a new child public key.

\subsubsection{Multiple Devices}
\label{sec:design:multi-device}
The client may interact with websites from multiple different devices and may subsequently want to issue \vcr{s} for one or more of these sessions.

By design, \sys{} allows clients to use any number of devices with a single trusted device.
Specifically, the master private key is used to generate a new \emph{device public key} which is stored on each of the client's devices.
The device public key can in turn be used for generating all other \vcr public keys needed on the device.
Note that even though the device public keys are derived from the same master private key, they are unlinkable and thus cannot be used by websites to link together sessions from the client's different devices.

\subsubsection{Secure Key Management}
\label{subsubsec:private_key_storage}
\vcr{} private keys for each session must be stored in a secure environment, access-controlled by the client. 
Leakage of private keys would allow an adversary to issue \vcr{s} for the client's sessions, thus giving them the ability to learn, modify, or delete, potentially sensitive information.

Our use of derivable asymmetric keys reduces the number of private keys that need to be stored securely to just one -- the master private key.
Another benefit of using derivable asymmetric keys is that the master private key can be stored offline.
This is because the master private key is only needed when generating a new device public key (i.e., when enrolling a new device) or creating \vcr{s}, which are expected to be relatively infrequent operations.

This feature provides \sys{} significant flexibility in terms of how the master private key is stored, in order to accommodate different levels of security.
For example, at one end of the spectrum, clients with low security requirements can simply store their keys on any device they trust, e.g., a phone or laptop.
Clients with higher security requirements can store their keys in hardware-backed keystores, such as the Android Keystore~\cite{androidkeystore} or Apple Secure Enclave~\cite{applesecureenclave}.
On PCs, clients could make use of hardware-enforced enclaves, such as Intel SGX~\cite{intelsgx} or Windows Virtualization-base security (VBS), to protect the keys.
At the top end of the spectrum, clients with the highest security requirements could store their keys in hardware security devices (e.g., YubiKey~\cite{yubikey}, Solokey~\cite{solokeys}, or Ledger~\cite{ledger}).
These clients may also enforce additional physical security controls, such as keeping the hardware security device in a locked safe until it is needed.
Clients can also make back-up copies of their master private keys to allow recovery if the trusted device fails.

We emphasize that a separate trusted device is {\em recommended} and not mandated. %
The only requirement for the trusted device is that it can derive child private keys and create signatures.
Section~\ref{sec:implementation} shows that these requirements can be met even by resource-constrained hardware security devices.

\subsubsection{Long-Term Storage}
\label{sec:design:storage}
A \vcr{} may be submitted long (possibly, years) after the corresponding session ends.
This typically requires clients to store state information for numerous sessions in a secure and highly available manner. 
Naturally, the amount of state per session must be minimal.

In \sys{} neither the cookies themselves nor the \wrapper{s} can be used to issue \vcr{s} without a signature from the client's master private key.
Therefore, the security requirements for the \emph{storage} of cookies and \wrapper{s} are minimal -- the integrity and availability of the cookies and \wrapper{s} must be maintained, but these pieces of information do not need to be kept confidential (assuming the cookies themselves have expired and are no longer useful, e.g., for authentication).

This opens up potential new business opportunities for third-party \emph{cookie storage} providers to offer a service for 
safely storing cookies and \wrapper{s} on clients' behalf. This service can take care of all cookie and wrapper management as well as
provide API endpoints to their customers. Note that security requirements for, and trust burden on, such services would be significantly 
higher if cookies alone were sufficient to issue \vcr{s}. Also, third-party cookie storage is not a requirement; 
clients who are uncomfortable with third-part providers storing their cookies can store them on their local devices.
Clients may also use their preferred cloud-storage service, e.g., OneDrive. %

\subsubsection{Broad Application Support}
\label{sec:design:multi-application}
Many non-browser applications also communicate with application-specific servers, which, similarly to Web servers, collect consumer data.
\sys{} is sufficiently flexible to be used in these applications as well.
This is achieved through the design pattern of using \wrapper{s} instead of directly modifying the session identifiers.
For example, if a non-browser application uses a different form of session identifier (i.e., not a Web cookie), \sys{} can still be used since the \wrapper{} can be used to cryptographically bind the client's \vcr{} public key to any type of session identifier.

\subsection{Overall \sys Design} \label{subsec:realizing_sys_design}
\label{sec:design:final}
\begin{figure*}[t]
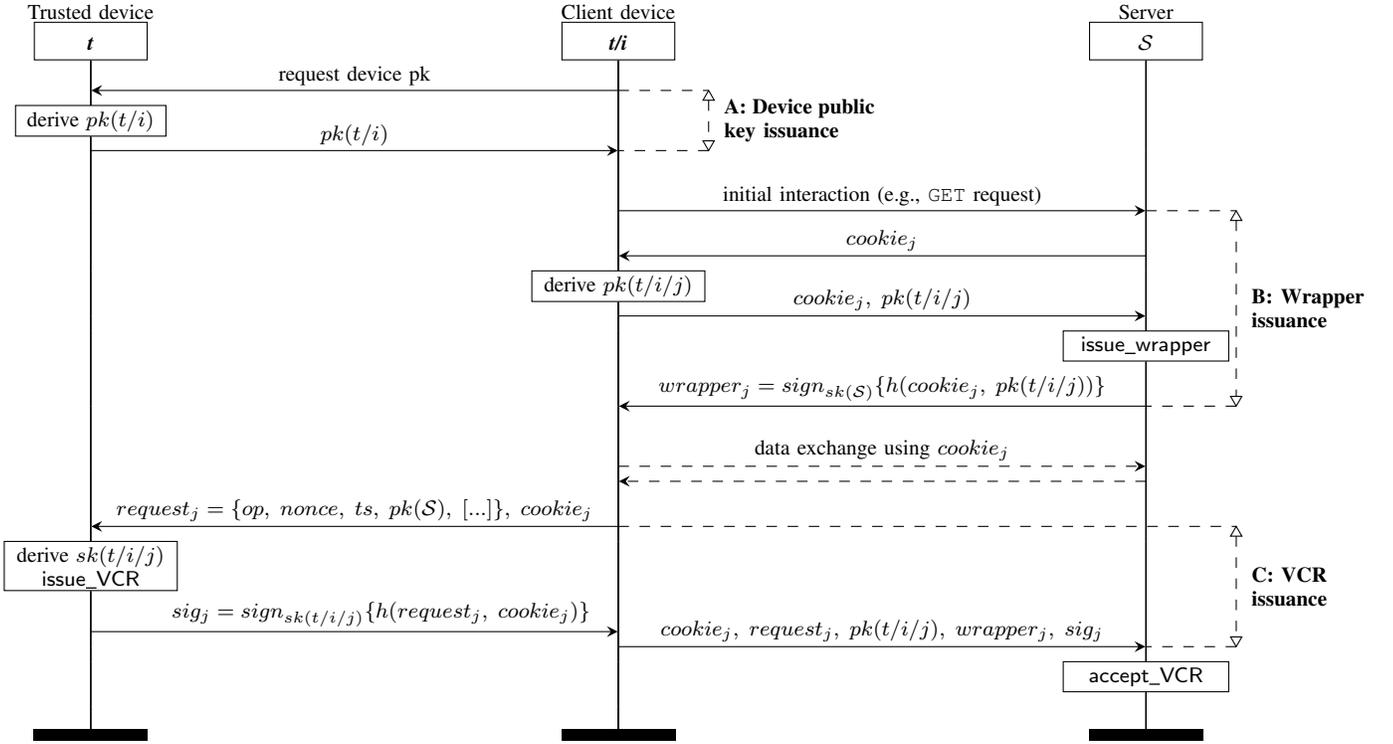

	\centering
	\begin{msc}[small values, draw frame=none, msc keyword=, head top distance=5mm, foot distance=0mm, instance distance=55mm, environment distance=10mm, level height=2mm, arrow scale=2.0, instance width=15mm, left inline overlap=15mm, right inline overlap=15mm, /tikz/font=\footnotesize, /tikz/line width=0.4pt, /msc/line width=0.4pt, label distance=0.4ex, action height=4mm]{ }
	
		\declinst{t}{Trusted device}{\textbf{\emph{t}}}
		\declinst{c}{Client device}{\textbf{\emph{t/i}}}
		\declinst{s}{Server}{$\mathcal{S}$}
		
		\measure[side=right, measure distance=12mm, label distance=2mm]{\parbox{20mm}{\textbf{A: Device public\\ key issuance}}}{c}{c}[4]
		\mess{request device pk}{c}{t}
		\nextlevel
		\action[action width=20mm]{derive $pk(t/i)$}{t}
		\nextlevel[3]
		\mess{$pk(t/i)$}{t}{c}
	
		\nextlevel[4]
	
		\measure[side=right, measure distance=12mm, label distance=2mm]{\parbox{15mm}{\textbf{B: \Wrapper\\ issuance}}}{s}{s}[13]
		\mess{initial interaction (e.g., \texttt{GET} request)}{c}{s}
		\nextlevel[3]
		\mess{$cookie_j$}{s}{c}
		\nextlevel
		\action[action width=23mm]{derive $pk(t/i/j)$}{c}
		\nextlevel[3]
		\mess{$cookie_j,\; pk(t/i/j)$}{c}{s}
		\nextlevel
		\action[action width=22mm]{$\mathsf{issue\_\wrapper}$}{s}
		\nextlevel[5]
		\mess{$wrapper_j = sign_{sk(\mathcal{S})}\{h(cookie_j,\; pk(t/i/j))\}$}{s}{c}
		\nextlevel[4]
		\mess*{data exchange using $cookie_j$}{c}{s}
		\nextlevel
		\mess*{}{s}{c}
		\nextlevel[3]
	
		\measure[side=right, measure distance=12mm, label distance=2mm]{\parbox{10mm}{\textbf{C: \vcr\\ issuance}}}{c}{s}[8]
		\mess{$request_j = \{op,\; nonce,\; ts,\;pk(\mathcal{S}),\; [...]\},\; cookie_j$}{c}{t}
		\nextlevel
		\action[action width=23mm]{derive $sk(t/i/j)$\\ $\mathsf{issue\_\vcr}$}{t}
		\nextlevel[6]
		\mess[pos=0.55]{$sig_{j} = sign_{sk(t/i/j)}\{h(request_j,\; cookie_j)\}$}{t}{c}
		\nextlevel
		\mess{$cookie_j,\; request_j,\; pk(t/i/j),\; wrapper_j,\; sig_{j}$}{c}{s}
		\nextlevel
		\action[action width=22mm]{$\mathsf{accept\_\vcr}$}{s}
		\nextlevel[3]
	
	\end{msc}
	\caption{Protocol messages exchanged in \sys. $sign_k(m)$ denotes a cryptographic signature on message $m$ using key $k$, and $h(m)$ denotes a cryptographic hash of message $m$. The events $\mathsf{issue\_\wrapper}(j)$, $\mathsf{issue\_\vcr}(j)$, and $\mathsf{accept\_\vcr}(j)$ are used in the formal security analysis in Section~\ref{sec:evaluation}.}
	\label{fig:protocol}
	\end{figure*}

Bringing together the design concepts discussed in the preceding sections, this section presents the overall design of \sys.
The precise protocol messages exchanged between the various principals are shown in Figure~\ref{fig:protocol} and the various cryptographic keys are defined in Table~\ref{tab:key_list}.

\subsubsection{Setup and device provisioning}
The client first generates a master private key $sk(t)$ on a trusted consumer device $t$.
This key is then used to derive a device-specific public key for each of device the client will use for interacting with websites, e.g., $pk(t/i)$ for device $i$.
Device public keys are provisioned using a simple request-response protocol, as shown in Figure~\ref{fig:protocol}~(\textbf{A}).
The device public keys are stored within the respective devices for rapid future \vcr key generation.

\begin{table}[t] \footnotesize
	\caption{List of keys and their role in \sys}
	\label{tab:key_list}
	\begin{tabularx}{\columnwidth}{l X}
	\toprule
	\textbf{Key} & \textbf{Name and description} 
	\\\midrule
	$sk(t)$ & Master private key: generated and stored in trusted device.
	\\\midrule
	$pk(t/i)$ & Device $i$ public key: derived from master private key $sk(t)$ within the trusted device, and stored on device $i$.
	\\\midrule
	$pk(t/i/j)$ & \vcr{} $j$ public key: derived from device public key $pk(t/i)$ when requesting a new \wrapper.
	\\\midrule
	$sk(t/i/j)$ & \vcr $j$ private key: derived from master private key $sk(t)$ within the trusted device when issuing a \vcr{}.\\\midrule
	$sk(\mathcal{S})$ & \Wrapper signing key: long-term private key used by server when generating a \wrapper.
	\\\midrule
	$pk(\mathcal{S})$ & \Wrapper public key: long-term public key used by client when verifying a \wrapper. Obtained via standard PKI.
    \\\bottomrule
	\end{tabularx}
\end{table}

\subsubsection{Website Interaction} 
When initiating a session with a server, upon a client request, the server generates a unique client id and sends it to the client in the form of a cookie.
The client generates a new \vcr{} public key $pk(t/i/j)$ for session $j$, derived from the device public key. 
The client then sends this newly-generated key, along with the client id cookie, to the server as shown in Figure~\ref{fig:protocol}~(\textbf{B}).
After receiving the \vcr{} key, the server signs the hash of the \vcr{} key and cookie to generate a \wrapper.
Then it returns the \wrapper{} to the client, attesting to the association between the client id cookie and the \vcr{} key.
To check the integrity of the \wrapper, the client verifies the signature using the server \wrapper public key which can be obtained and verified the same way it obtains and verifies the server's TLS public key.

\subsubsection{\vcr Issuance}
\label{sec:design:vcr_issuance}

As shown in Figure~\ref{fig:protocol}~(\textbf{C}), the client generates a request, which consists of the specified operation (e.g., retrieve, modify, or delete), the current time, 
and any optional parameters.
The client uses the master private key $sk(t)$ on the trusted device to derive the respective signing key $sk(t/i/j)$ for device $i$ and session $j$.
The signing key $sk(t/i/j)$ is then used to sign a hash of the request and the associated cookie, which is returned to the client.
The client then sends this signature, along with the cookie, request, public key, and \wrapper{} to the server, constituting the \vcr.

Upon receiving a \vcr{}, the server first verifies its own signature on the \wrapper{} to confirm the authenticity of the \wrapper{}. 
The server then verifies the client's signature on the \vcr{} using the public key $pk(t/i/j)$ from the \wrapper{}.
If these checks succeed, the server accepts the \vcr and proceeds with the requested data operation.

To prevent an attacker replaying a valid \vcr{} to the server, the client includes a unique $nonce$ in the request, which is thus also included in the client's signature $sig_j$ in Figure~\ref{fig:protocol}.
Upon receiving this \vcr, the server checks that it has not already processed a \vcr containing that $nonce$.
The client also includes a timestamp $ts$ in the request, representing the time at which the \vcr was issued.
Each server defines its own recency threshold (e.g., 12~hours) and rejects any \vcr{s} that are older than this threshold.
This means that the server only has to store nonces for up to this threshold in order to check that new requests are unique.
This also ensures that an attacker cannot delay valid \vcr{s} arbitrarily (e.g., if the attacker were able to block a \vcr and then release it months later, this could have unintended consequences for consumers).

An alternative would be to use a challenge-response protocol where the server generates a challenge that the client must include in the signed \vcr{}.
This would avoid the need for a nonce and timestamp, but would increase the complexity of the system and could be abused to mount a denial of service attack against the server, similarly to a TCP SYN flooding attack (although well-known cookie-based countermeasures and application level solutions such as CAPTCHAs~\cite{von2003captcha} or rate-limiting mechanisms~\cite{272094} could also be used.).

\subsubsection{Device Unprovisioning}
\label{subsubsec:unprovisioning}
Finally, the full device lifecycle may require \emph{unprovisioning}, e.g., if the device is lost/stolen, or sold/recycled.
We separately consider the implications for a regular or trusted device.

\taggedpara{Regular Device}
Since such devices do not hold private \vcr{} keys, unprovisioning is straight-forward.
The client only has to unlink the old device from the respective trusted device to prevent any further \vcr{} issuance.
This can be done from the trusted device, even if the old device is lost/stolen. 
The client may wish to back-up or transfer any cookies and \wrapper{s} from the old device. 

\taggedpara{Trusted Device}
From an availability perspective, the client should be able to recover the master private key from a backup.
From a security perspective, the trusted device should ideally have some type of access control (e.g., using a PIN and/or a fingerprint) to protect the private key even from an adversary who has physical access to the device.
If the trusted device is being sold/recycled, the client should securely back-up or transfer the master private key to a new trusted device using techniques such as Presence Attestation~\cite{Zhang2017}.

\section{Implementation}\label{sec:implementation}
We now describe our implementation of \sys{}, which consists of: a browser extension, a trusted device implementation, and a modified web server.

\subsection{Browser Extension}\label{subsec:ext}
This component provides most of the client-side functionality on a regular consumer device.
Specifically, it manages \vcr{} public keys, implements client-side aspects of the \vcr{} flow, and includes a 
client-facing interface for controlling this flow.
It also handles communication between the browser and the trusted device (or application) that holds the master private key.
Although we chose to implement a proof-of-concept extension for Google Chrome, no (or only minor) mods would be required to get
it to work with any modern browser.

To realize derivable asymmetric keys, we used a mechanism proposed for hierarchical deterministic wallets, 
commonly known as Bitcoin Improvement Proposal 32~\cite{bip32}, or \texttt{BIP32}.
\texttt{BIP32} has the notion of an \emph{extended key}, with 256 bits of entropy (called \emph{chain code}) added to a normal public/private key-pair.
Extended keys can be used to derive one or more child keys, following the rule that private keys can be used to derive private or public keys, 
while public keys can only derive public keys.\footnote{\texttt{BIP32} also has the notion of \emph{hardened} vs.\ \emph{non-hardened} keys, though we only use the latter. The difference between these two is the algorithm used when deriving them from the parent key.}
\texttt{BIP32} is also well-suited for low-end devices, since it was designed for (resource-constrained) Bitcoin hardware wallets.
Section~\ref{subsec:trusted_device} describes our proof-of-concept of a resource-constrained trusted device.

The browser extension comprises a background and a pop-up scripts, both written in \texttt{JavaScript}, 
with additional \texttt{HTML} and \texttt{CSS} for the pop-up.
As no \texttt{JavaScript} version of several \texttt{Node.js} libraries (e.g., \texttt{crypto}, \texttt{BIP32}) was available, we used Browserify~\cite{browserify} to convert such libraries into \texttt{JavaScript} files that can be loaded by the browser.

\subsubsection{Background Script} \label{subsec:background_script}
This component houses most \vcr-side functionality.
It uses the browser's API (\texttt{chrome.webRequest.onHeadersReceived}) to scan \texttt{HTTP} response 
headers to detect which servers support \sys{}. If present, it parses the relevant \sys{} endpoints and the 
client id cookie from the headers.

Using the device public key, it derives a session-specific \vcr{} public key using a port of the official \texttt{BIP32} implementation~\cite{bip32code}.
The derivation path of a \vcr{} public key is in the form \texttt{t/i/j}, where \texttt{t} denotes the master private key, \texttt{i} the device ID, and \texttt{j} the total number of sessions created on the device.
In other words, \texttt{t/i} represents the device public key and \texttt{t/i/j} represents the child public key for the $j^{th}$ session. 

After deriving a \vcr{} public key, the background script either includes this in the next request header sent to the server, 
or sends a \texttt{POST} request to a separate server-defined \vcr{} \wrapper{} request endpoint (we implemented the latter). 
The server then responds with a newly generated \wrapper.

The background script then stores the server-returned \wrapper{} along with the key derivation path.
Since we generate a new one for each session, the number of stored \wrapper{s} may grow large, depending on how many new sessions the client establishes.
However, the storage overhead of the \wrapper{s} is not significant, as the number of \wrapper{s} is at most the same as the number of cookies the client must store (see Section~\ref{sec:storage_eval} for client-side storage evaluation).
To improve efficiency of searching for a client id cookie, we use a hashmap of client id 
cookies and their corresponding \wrapper{} information. We also store the URL of the 
website and the time of the visit alongside the \wrapper{} to assist clients when selecting a session during \vcr{} issuance.
The background script can store this data using any storage service.
Our implementation uses the local storage API (\texttt{chrome.storage.local}).
Other choices include cloud storage services or Google Chrome's synced storage API (\texttt{chrome.storage.sync}),
which would allow clients to synchronize \sys{} data between different devices.\footnote{Unfortunately, we found that Chrome 
synced storage currently imposes a limit on the amount of stored data.}

Note that all the above operations are performed asynchronously, in the background. Thus, 
the client does not experience any additional latency in loading the page.

\subsubsection{Pop-up}
\begin{figure}[t]
	\centering
	\fbox{\includegraphics[width=0.7\linewidth]{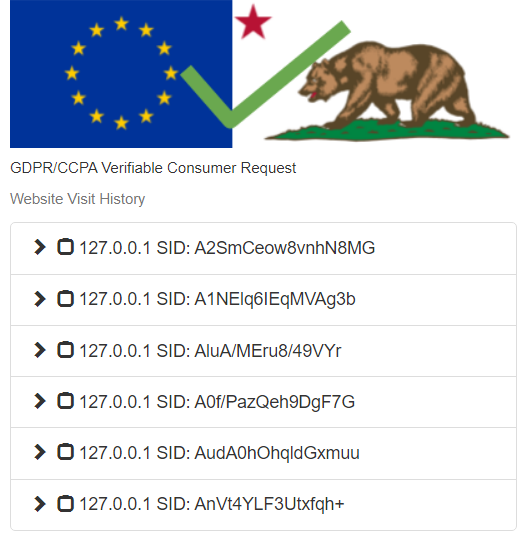}}
	\caption{\sys{} browser extension pop-up displaying multiple sessions. 
	\texttt{SID} is the first few bytes of \vcr{} public key.}
	\label{vcr_list}
\end{figure}

\begin{figure}[t]
	\centering
	\fbox{\includegraphics[width=0.7\linewidth]{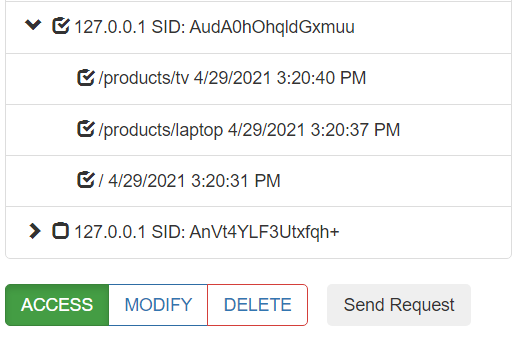}}
	\caption{\sys browser extension pop-up displaying the history of visits in each session. 
	Clients can select which session and which type of \vcr{} (\texttt{ACCESS/MODIFY/DELETE}) they wish to generate.}
	\label{vcr_history}
\end{figure}

The extension pop-up is a small web page that appears when the client clicks on the extension icon next to the URL bar. 
As shown in Figures~\ref{vcr_list} and~\ref{vcr_history}, it displays session information for a \vcr{} key along with the history 
of web links visited when the session was active. 
It also displays the types of \vcr{s} (access, modify, and delete) that the client can submit.

To issue a \vcr{}, the client first chooses the session(s) and the type of request.
Next, the pop-up script prepares a request for signing.
It includes a timestamp in the signature to prevent replay attacks, due to its simple design.
The resulting client request is then passed to a Python native application~\cite{nativemessagingprotocol} 
which relays it to the trusted device (see Section~\ref{subsec:trusted_device}) for signing.
Once signed using the \vcr{} private key, the request is retrieved by the pop-up (using the background script)
through the native application, and the \vcr{} is sent to the server's \vcr{} verification endpoint.
Finally, the server's response is displayed in the client's browser.

\subsection{Native Messaging Application} \label{subsec:native_message_app}
To simulate a client-controlled trusted device, we created a \texttt{Node.js}~\cite{nodejs} application that holds the master private key.
Alike the trusted device, this application signs \vcr{s} received from the background script using private keys derived from 
the master private key with the provided key derivation path, e.g., \texttt{t/0/1}.
Communication between this application and the browser is done via the native messaging protocol~\cite{nativemessagingprotocol}.
We use the \texttt{native-messaging} package~\cite{nativemessagingjs} to support both Firefox and Chrome.
As mentioned in Section~\ref{subsubsec:private_key_storage}, the key storage mechanism must 
support derivable asymmetric key operations (e.g., using \texttt{BIP32}).
There are publicly available implementations of \texttt{BIP32} in different languages, including: 
JavaScript~\cite{bip32code}, Golang~\cite{bip32golang}, Python~\cite{bip32python}, Java~\cite{bip32java} 
and C~\cite{trezor-github}.

\subsection{Trusted Device} \label{subsec:trusted_device}

\begin{figure}[t]
	\begin{minipage}[b]{0.45\linewidth}
	\centering
	    \centering
	    \includegraphics[width=1\linewidth]{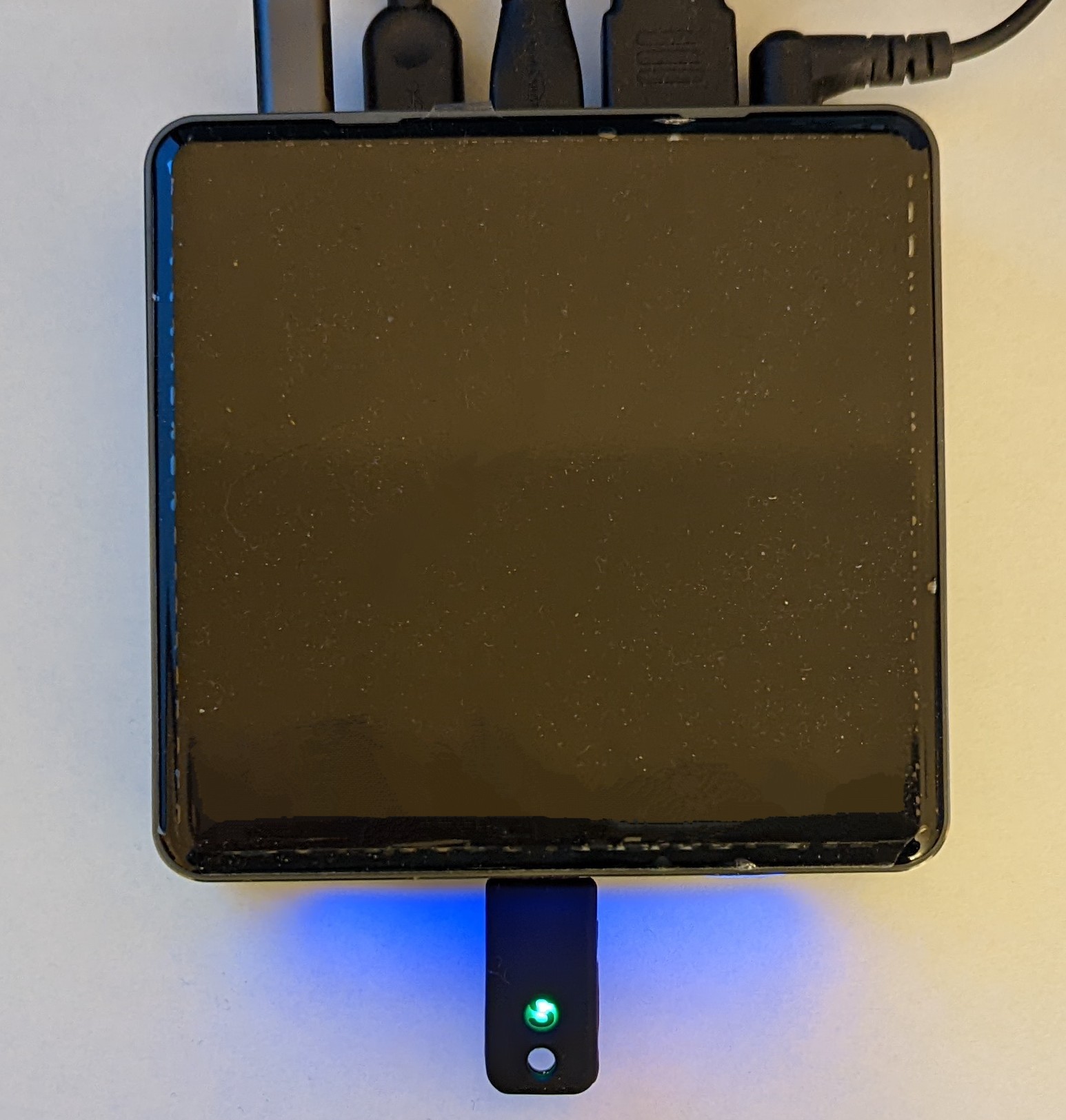}
		\caption{Solokey hardware security token in use.}
		\label{fig:solokey_use}
	\end{minipage}\hfill
	\begin{minipage}[b]{0.45\linewidth}
	    \centering
	    \includegraphics[width=1\linewidth]{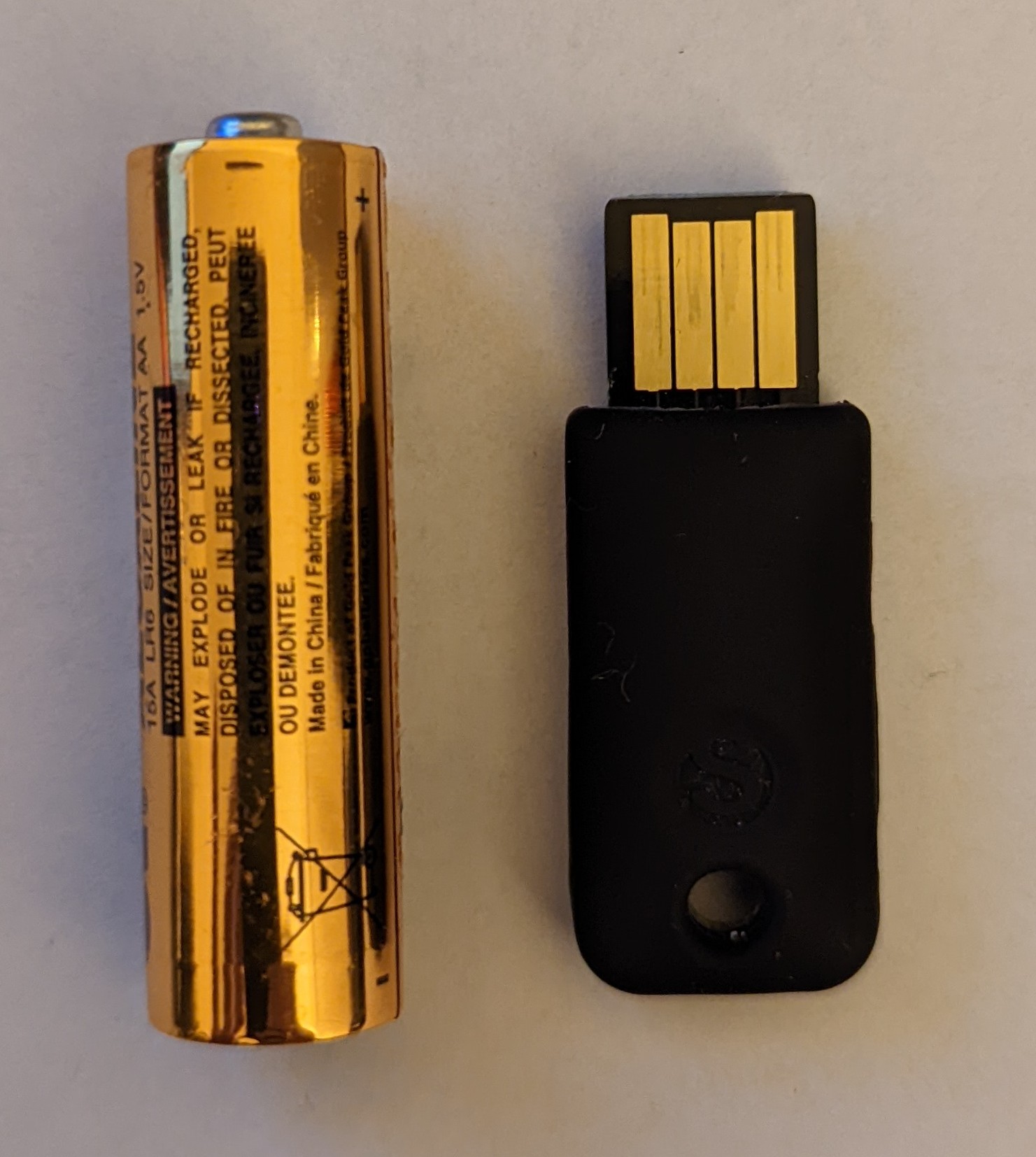}
		\caption{Solokey token compared to a standard AA battery.}
		\label{fig:solokey_comparison}
	\end{minipage}
\end{figure}

We implemented a proof-of-concept trusted device using the Solokey Hacker~\cite{solokeys}, a security token with open source software and hardware.
Solokey includes an STM32L432KC microprocessor with an Arm Cortex-M4 MCU (80MHz), 64~kB of RAM, 256~kB of 
flash memory, a true random number generator (TRNG), and a physical button for presence attestation, as shown in
Figures~\ref{fig:solokey_use} and~\ref{fig:solokey_comparison}. We extended the FIDO2 Client to Authenticator Protocol 
(CTAP) API on the device with the following three calls:
\begin{compactitem}
	\item \texttt{KEYGEN}: Generates a master private key using the TRNG. This key never leaves Solokey.
	\item \texttt{DEVKEY}: Takes a key derivation path (\texttt{t/i}) and outputs a generated device public key (see Figure~\ref{fig:protocol}~(\textbf{A})).
	\item \texttt{VCRGEN}: Takes a key derivation path (\texttt{t/i/j}) and a consumer request (in the form of a cryptographic hash), and outputs a 
	signed \vcr{} (see Figure~\ref{fig:protocol}~(\textbf{C})).
\end{compactitem}
All three trusted device API calls require human confirmation achieved by the client pressing the physical button on the Solokey.
For key derivation and signing we used the \texttt{BIP32} implementation from Trezor~\cite{trezor-github} 
due to its popularity and suitability for embedded devices.
The native messaging application calls the individual API according to the command it receives from the background script.
Once the client confirms the action by pressing the button and the native messaging application receives a response from 
Solokey, the application relays the response back to the background script.

To protect against tampering with the consumer request by client-side malware, the trusted device could also display information 
about the \vcr{} that is about to be generated, including the request type, the website URL, and the timestamp.
In our prototype, this could be as simple as changing LED color on the Solokey.
Displaying more detailed information about the request may require trusted devices to 
have a dedicated, human-perceivable output means, e.g., a display.

\subsection{\sys{}-enabled Web Server}\label{subsec:sv}
We implemented a proof-of-concept \sys{}-enabled \texttt{HTTP} server using the Express Node.js web framework~\cite{nodeexpress}.
It uses \texttt{HTTP} sessions and indexes data collected during each session using session cookies.
For signing, it uses ECDSA with curve \texttt{secp256k1}~\cite{nodesecp256k1}, although any secure signature 
scheme can be used.

When the client first visits a web page hosted on our server, the latter creates an \texttt{HTTP} cookie that includes 
a client id (uuid~\cite{nodeuuid}).Hereafter, all data collected by the server about this client is associated with 
the client's unique id.\footnote{\sys can also use any existing client identifier cookie scheme -- such as Google 
Analytics' \texttt{\_ga} and \texttt{\_gid} cookies~\cite{ga}.}
In response to the initial client request, the server notifies the client that it supports \sys{}.

The server provides the client with an \texttt{HTTP} endpoint for obtaining a \wrapper.
The client's browser extension sends the client id cookie set by the server and a freshly-generated \vcr{} public key to this endpoint, and the server then generates a \wrapper{} that cryptographically binds these two pieces of information. 
This endpoint can be configured to only issue \wrapper{s} for a short duration after the cookie was issued, to avoid adversaries attempting to bind their own public keys to arbitrary cookies.
Alternatively, the client can provide a freshly-generated \vcr{} public key in the next \texttt{HTTP} request, and the server can issue the \wrapper in the next response.

The server also provides the client with a \vcr{} endpoint, to which the client can later submit \vcr{s} for this website.
As discussed in Section~\ref{sec:design:final}, the client sends a \wrapper{} and a signed \vcr{}, which the server then verifies.
To support requested \vcr{} actions, the consumer's request may include metadata specific to the requested action.
For instance, for data access requests, metadata may include an encryption key to be 
used by the server to encrypt data to be returned to the client.
\section{Evaluation} \label{sec:evaluation}
We now evaluate security of \sys as well as its latency, data transfer, and storage requirements.

\subsection{Security Analysis} \label{subsec:security_analysis}

To evaluate security of \sys{}, we defined a formal specification of the protocol using the Tamarin prover~\cite{meier2013tamarin,tamarin-website}.
The full specification is provided in Listing~\ref{lst:tamarin} in the Appendix. %
In Tamarin, a protocol $P$ is modelled as a set of labelled transition rules operating on \emph{facts}.
A transition consumes linear facts from state $s_{i-1}$, generates new facts for state $s_{i}$, and labels the transition as $a$.
An execution of protocol $P$ is a finite sequence of states and transition labels $(s_0, a_1, s_1,\ldots, a_n, s_n)$ such that $s_0 = \emptyset$ and $s_{i-1}\xrightarrow{{a_i}}s_i$ for $1\leq i\leq n$.
The sequence of transition labels $(a_1,\dots,a_n)$ is a \emph{trace} of $P$ and the set of all traces is denoted $traces(P)$.
Security properties are specified using first-order logic formulas on traces.

\taggedpara{Unforgeability} Using this specification, we formally define the security property of \emph{unforgeability} (see Section~\ref{sec:system_and_threat_models}), for both \wrapper{s} and \vcr{s}.
As shown in Figure~\ref{fig:protocol}, let $\mathsf{issue\_\wrapper}(s, o, pk)$ denote server $s$ issuing a \wrapper for cookie $o$ and public key $pk$ (corresponding to private key $sk$);
let $\mathsf{issue\_\vcr}(c, o, pk, r)$ denote client $c$ issuing a \vcr for request $r$, cookie $o$, and public key $pk$;
and let $\mathsf{accept\_\vcr}(s, o, pk, r)$ denote server $s$ accepting a \vcr for request $r$, cookie $o$, and public key $pk$.

\begin{definition}[\Wrapper unforgeability]\label{eq:wrapper_unforgeability}
A protocol $P$ satisfies the property of \emph{\wrapper unforgeability} if for every $\alpha \in traces(P)$:
\begin{align*}
	\forall s, o, pk, r, j.\ &\mathsf{accept\_\vcr}(s, o, pk, r)\in \alpha_j\implies\\
	&\exists i.\ \mathsf{issue\_\wrapper}(s, o, pk)\in \alpha_i \wedge i < j
\end{align*}
\end{definition}

\begin{definition}[\vcr unforgeability]\label{eq:vcr_unforgeability}
A protocol $P$ satisfies the property of \emph{\vcr unforgeability} if for every $\alpha \in traces(P)$:
\begin{align*}
	\forall s, c, o, pk, r, j.\ &\mathsf{accept\_\vcr}(s, o, pk, r)\in \alpha_j\implies\\
	&\exists i.\ \mathsf{issue\_\vcr}(c, o, pk, r)\in \alpha_i \wedge i < j
\end{align*}
\end{definition}

The Tamarin prover verifies that, in the protocol as specified, both of these properties hold for an unbounded number of protocol runs (the strongest possible result).
Since \wrapper{s} and \vcr{s} are ultimately verified by the server, Definition~\ref{eq:wrapper_unforgeability} requires that, whenever a server accepts a \vcr, that server must have issued a corresponding \wrapper at some prior time point.
Similarly, Definition~\ref{eq:vcr_unforgeability} requires that, whenever a server accepts a \vcr, there must exist a client that issued that \vcr at some prior time point.
Intuitively, these properties show that the adversary cannot forge a valid \wrapper or \vcr for a given $pk$.
Since the corresponding private key is only known to the client, we conclude that the unforgeability property is satisfied.

\taggedpara{Replay resistance} To prevent an attacker obtaining a genuine \vcr (e.g., by eavesdropping) and later 
replaying it to the server, we formally define an the security property of \emph{replay resistance} (see Section~\ref{sec:system_and_threat_models}) for \vcr{s}.

\begin{definition}[Replay resistance]\label{eq:replay_resistance}
A protocol $P$ satisfies the property of \emph{replay resistance} if for every $\alpha \in traces(P)$:
\begin{align*}
	\forall s, o, pk, r, j.\	&\mathsf{accept\_\vcr}(s, o, pk, r)\in \alpha_i\; \wedge\\
								&\mathsf{accept\_\vcr}(s, o, pk, r)\in \alpha_j\implies i = j
\end{align*}
\end{definition}

The Tamarin prover verifies that this property holds for an unbounded number of protocol runs.
Definition~\ref{eq:replay_resistance} states that, if there are two trace events in which a server accepts the same \vcr, these must be the same event.
Since Tamarin does not model time-based properties, the formal model uses only a nonce in the \vcr to check for uniqueness.
In practice, the client would also include a timestamp in the \vcr, as described in Section~\ref{sec:design:vcr_issuance}.

\taggedpara{Consumer/Device/Request Linking} To protect clients' privacy, an honest-but-curious server should be unable to link a \vcr to a specific client, or to link multiple \vcr{s} to the same client (see Section~\ref{sec:system_and_threat_models}).
This requirement ensures that the use of \sys does not reveal any additional information to the server about potential links between users, devices, and sessions (e.g., if a single user is using multiple devices).

Since unlinkability is not a trace property, we cannot use Tamarin to model or verify this property.
Instead we follow an existing approach for reasoning about unlinkability~\cite{veeningen2011formal,veeningen2010modeling,paverd2014modelling} and show that the messages sent by the client to the server do not contain any information that could be used by a server to link \vcr{s} to clients or to other \vcr{s}.

As shown in Figure~\ref{fig:protocol}, the only new pieces of information provided by the client (which the server does not already know) are (1) the $request$, (2) the \vcr public key ($pk(t/i/j)$), and (3) the client's signature.
The $request$ does not contain any information that uniquely identifies the client or allows it to be linked to other requests.
Formally, given any two requests, a server would not be able to distinguish whether or not they were issued by the same client.
\texttt{BIP32} guarantees that derived public keys are unlinkable to each other and to their parent keys.
This ensures that neither the \vcr public key nor the signature created using the corresponding private key are linkable.\footnote{By design, a \vcr can be linked to the corresponding \wrapper{} -- indeed the latter is included in the former.}
Formally, given any two derived public keys or signatures, a server would not be able to distinguish whether or not they were issued by the same client.
We can therefore conclude that the protocol satisfies the \emph{unlinkability} property.

Of course, if the device public key is leaked from the client's device, different \vcr{} could be linkable.
However, even in this case, \sys is no worse than the current use of web cookies,
since an attacker that can steal a device public key could also steal cookies from the victim's 
device and use these to link/track the victim's sessions and \vcr{s}.

Although \sys provides unlinkability by design, the nature of how \vcr{s} are submitted may 
point servers in the direction of clients. For instance, by observing metadata such as IP addresses of
different \vcr{} requests, a server might link them to the same client. One possible mitigation is to 
use anonymity networks (e.g., Tor) and avoid issuing \vcr{} \emph{bouquets} whereby 
multiple requests are submitted through the same connection. Random delays between \vcr{s} can be used to prevent timing-based correlation.

\taggedpara{Public Key Injection} The adversary might attempt to replace the client \vcr{} 
public key with its own public key when obtaining a \wrapper{} for a cookie. 
This can occur if there is either: \emph{(1)} an active network-level adversary, and/or \emph{(2)} malware on client device.
For (1), this is mitigated by the use of secure communication channels (e.g. TLS) or by simply having  
\sys{} browser extension compare the public key it sent with the public key in the returned \wrapper{}.
In contrast, (2) is difficult to defend against.
A malware in full control of the client device, can replace the client public key with its own during the \wrapper{} request.
Even if the consumer attempts to verify the public key in the returned \wrapper{}, the malware can subvert this check.
The only way to prevent this is to verify \wrapper{s} in an environment isolated from client-side malware,
e.g., a Trusted Execution Environment (TEE).
Of course this approach would also require securely sharing the public key to be verified with the TEE and displaying the verification result to the user without malware interference.
Overall, we consider (2) to be out of scope since malware in total control of a client device already has access to any data that could be collected by the server about the client.

\taggedpara{Client-Side Malware}
Malware on the client's device might conduct unauthorized operations on client data.
This might be possible either via: \emph{(1)} replay attacks, or \emph{(2)} by generating \vcr{s} without the
owner's consent. %
This issue highlights one important difference between using asymmetric tokens and symmetric tokens for \vcr{s}.
The former allows generation of one \vcr{} per client authorization. In contrast, symmetric tokens, even if encrypted and 
decrypted on demand with client's approval, need to be available in plaintext at some point in order to be sent to the servers.
Client-side malware can use these exposed tokens to generate future \vcr{s}.
Also, using symmetric tokens allows malware to access data from \emph{before} its infection period. For example, 
assume that malware infects the client's device at time $t$.
It can access pre-stored tokens and learn data generated \emph{prior} to $t$. 
Since we can prevent such attacks via asymmetric tokens, \sys{} provides better security and 
privacy compared to current symmetric token-based systems.

\taggedpara{Key Leakage} An attacker might exploit weaknesses in \texttt{BIP32} to learn the private key.
One well-known weakness of \texttt{BIP32} is that knowledge of a parent extended public key as well as of 
any non-hardened child private key (descended from that parent public key) can leak the parent extended private key~\cite{bip32}. 
However, in \sys, the non-hardened child private key is generated within, and never leaves, the trusted device.
Therefore, the attacker must compromise the trusted device to obtain the non-hardened child private key, 
which we consider to be infeasible, per Section~\ref{sec:system_and_threat_models}.

\subsection{Latency Analysis}
\label{subsec:latency_analysis}
Most operations in \sys result in no user-perceptible latency because they occur asynchronously with normal web browsing.
Nevertheless, we discuss them to quantify computational costs of \sys{}.
The only user-perceivable latency occurs when a \vcr{} is issued, which is expected to be an infrequent operation.
For these experiments, we used as the client device an Intel NUC with an Intel Core i5-7260U 2.20GHz 
quad-core CPU with 32.0 GB of RAM running Ubuntu 18.04 LTS, with Chrome version 97.0.4692.71 64-bit official build. 
Unless otherwise stated, all results are averages over 10 runs, with storage left unchanged between runs. 
All data was in local storage and results may vary depending on the underlying storage technology, 
e.g., memory vs.\ hard-drives vs.\ cloud-hosted databases.

\taggedpara{Obtaining a \Wrapper{}}
We divide the process of obtaining a \wrapper{} into the following four phases: 
\newcommand{\prewrapper}{Key Derivation\xspace}
\newcommand{\prewrappertime}{24.6~ms}
\newcommand{\wrapperprep}{\Wrapper Generation\xspace}
\newcommand{\wrapperpreptime}{0.4~ms}
\newcommand{\postwrapper}{\Wrapper Verification\xspace}
\newcommand{\postwrappertime}{18.8~ms}
\newcommand{\storewrapper}{\Wrapper Storage\xspace}
\newcommand{\storewrappertime}{6.5~ms}
\begin{compactenum}
	\item \textit{\prewrapper}:
	When an unknown client makes a request, the server returns a cookie and the \vcr{} endpoints.
	The client parses these endpoints, derives a \vcr{} public key, and prepares a \wrapper{} request.
	\item \textit{\wrapperprep}:
	The server generates a \wrapper{} using the client-provided \vcr{} public key and cookie.
	\item \textit{\postwrapper}:
	After receiving the \wrapper{} from the server, the client verifies the \wrapper{} using the server's 
	public key and confirms that the \wrapper{} associates the correct \vcr{} public key and cookie.
	\item \textit{\storewrapper}:
	The client saves the \wrapper{} along with the public key derivation path and endpoints.
\end{compactenum}

\begin{table}[t] \small
\centering
\caption{Latency Results for \sys \Wrapper{s}.}
\label{wrapperLatencyTable} 
\begin{tabularx}{\columnwidth}{X X X X}
\toprule
\textbf{Key Derivation} & \textbf{\Wrapper{} Generation} & \textbf{\Wrapper{} Verification} & \textbf{\Wrapper{} Storage} \\ 
\midrule
\prewrappertime & \wrapperpreptime & \postwrappertime & \storewrappertime \\
\bottomrule
\end{tabularx}
\end{table}
Table~\ref{wrapperLatencyTable} shows the average time of each phase.
These measurements exclude network latency, as this will vary depending on the locations of the client and server.
\postwrapper takes the longest as it includes a signature verification.
\wrapperprep is noticeably faster than \prewrapper and \postwrapper 
since the former is performed by a native application and the latter two run in the browser extension.

\taggedpara{Issuing a \vcr}
We divide the process of issuing a \vcr into two steps:

\newcommand{\prevcr}{\vcr Generation\xspace}
\newcommand{\prevcrtime}{1357.4~ms\xspace}
\newcommand{\postvcr}{\vcr Verification\xspace}
\newcommand{\postvcrtime}{1.5~ms\xspace}

\textit{\prevcr}: When a client selects a session and a \vcr type, the browser extension prepares a request to be signed by 
the trusted device and a key derivation path. Both are sent to the trusted device via the native messaging application.
Next, the trusted device signs the overall request using the private key corresponding to the derived public key.
Finally, the signature is then returned to the extension.
The above steps in total take on average \prevcrtime using a modified Solokey Hacker as the trusted device.

\textit{\postvcr}: The server receives the \vcr and verifies the \wrapper.
Also, it extracts the \vcr public key for this session from the \wrapper and verifies the overall \vcr using the \vcr public key.
This step takes on average \postvcrtime.

\begin{table}[t] \small
\centering
\caption{\vcr Latency Results.}
\label{vcrLatencyTable}
\begin{tabularx}{\columnwidth}{X X X}
\toprule
          & \textbf{\vcr Generation} & \textbf{\vcr Verification} \\
\midrule
\vcr Flow & \prevcrtime     & \postvcrtime      \\
\bottomrule
\end{tabularx}
\end{table}

Table~\ref{vcrLatencyTable} shows latency results (excluding network latency).
Similar to Table~\ref{wrapperLatencyTable}, a signature generation operation performed by the native messaging application 
takes longer, compared to a standalone server. This is due to data passing delay between pop-up and background scripts, 
as well as the native messaging protocol between the application and the browser.
Based on these results, server's cost to verify \vcr{s} is minimal.

\newcommand{\keygentime}{332~ms\xspace}
\newcommand{\devkeytime}{724~ms\xspace}
\newcommand{\vcrgentime}{1348~ms\xspace}

\taggedpara{Trusted Device Latency}
Finally, we benchmarked the latency of key-generation operations performed by the trusted device.
Generation of the master private key takes \keygentime and generation of a device key takes \devkeytime, which are both
reasonable for these types of operations. As discussed in Section~\ref{sec:implementation}, we use a modified Solokey 
Hacker as an example of a trusted device, while noting that this resource-constrained hardware token is likely to be the 
slowest type of a trusted device. We emphasize that these are very infrequent operations, taking place once per trusted 
device, and once per new client device respectively.

\subsection{Data Transfer Analysis}
We measured the amount of data transferred to obtain \wrapper{s} and issue \vcr{s}.
For demonstration purposes, our server kept the visit history for each client, which was returned to the client upon successful \vcr{} verification.
The setting was the same as for latency analysis, and we used the browser's debugging console to measure the amount of data exchanged between the browser and server.

The client first sent an \texttt{HTTP GET} request to the server.
After receiving the \vcr{} endpoints and a cookie, the client sent a \texttt{POST} request to the \wrapper{} request endpoint.
The client then generated and sent a \vcr{} to the server requesting to access the data, and received the visit history with a single entry.
This request included: \wrapper{}, \vcr{} public key, and signature on the request.
Table~\ref{tab:optimized_bandwidth} shows the \texttt{HTTP} header and payload sizes transmitted between the client and server.
For both obtaining a \wrapper{} and issuing a \vcr{}, the amount of data exchanged was a fraction of a typical web interaction.

\newcommand{\reqwrapperload}{\shortstack{0.72 kB\\(0.62 + 0.10)}}
\newcommand{\reswrapperload}{\shortstack{0.38 kB\\(0.23 + 0.15)}}
\newcommand{\totalwrapperload}{1.10 kB}

\newcommand{\reqvcrload}{\shortstack{0.99 kB\\(0.68 + 0.31)}}
\newcommand{\resvcrload}{\shortstack{0.28 kB\\(0.23 + 0.05)}}
\newcommand{\totalvcrload}{1.27 kB}

\begin{table}[t] \small
	\centering
	\caption{Data transfer in kB (HTTP header + payload).}
	\label{tab:optimized_bandwidth}
	\begin{tabularx}{\columnwidth}{X c c c}
	\toprule
					& \textbf{Request}         & \textbf{Response} 	    & \textbf{Total} \\
	\midrule
	Obtain \Wrapper & \reqwrapperload & \reswrapperload & \totalwrapperload \\
	Issue \vcr      & \reqvcrload     & \resvcrload     & \totalvcrload     \\
	\bottomrule
	\end{tabularx}
\end{table}

\subsection{Storage Analysis}
\label{sec:storage_eval}
For the client-side data storage evaluation, we measured client-side data storage requirements using the \texttt{chrome.storage.sync.getBytesUsed} function.
The date was represented using the UNIX standard and the only the URL path was stored in the history section.

We first measured the minimum storage for a client with no visit history: it stores server endpoints, \vcr{} and server public keys, plus metadata used to issue \vcr{s}, requiring 0.38 kB in total.
We then measured storage for a client who visited a particular URL 100 times, where \sys{} stored the details of each visit.
This required 5.06~kB. Although storage increases linearly with the number of visited more web pages, the gradient is small 
and the overall magnitude is similar to that of a typical web browser history.

We can extrapolate from the results above to estimate the amount of storage required by a typical client to store all \wrapper{s} generated over a long period of time.
Crichton et al.~\cite{crichton2021browse} recently found that, on average, a user visits $163$ distinct web pages per day.
Note that ``distinct web page'' refers to a unique URL and not necessarily to a unique domain, i.e., the number of distinct domains may be smaller.
This study does not report the fraction of web pages where a user has an account.
We make the following conservative assumptions: (i) the $163$ web pages correspond to distinct domains; (2) the user has no accounts on any of these web pages; and (3) the user stores all \wrapper{s} for one year before issuing a \vcr request.\footnote{Note that this analysis is purely illustrative and that the storage requirements could increase or decrease if these assumptions change.}
Under these assumptions, \sys would require $0.38$~kB for each of the $163$ web pages for $365$ days, resulting in a total storage requirement of $22.61$~MB.

\subsection{Deployability Analysis} \label{subsec:deployability}
From the server perspective, main changes are: (1) create and maintain a public/private key-pair for generating \wrapper{s}, and (2) create and maintain relevant endpoints.
By design, the server does not have to change how it assigns identifiers to clients or uses cookies.
The integration of \sys into existing servers can be further simplified by releasing \sys{} modules for popular server frameworks.
From the server's perspective, \sys{} is a cheaper and simpler approach to comply with data protection regulations, compared to existing third-party identity verification services.

From the client's perspective, \sys requires: (1) generate a master private key on a trusted device, and (2) install the browser extension and native messaging application on other devices.
These software packages could be made available via the popular app stores.
Once installed, \sys operates transparently to the client, and does not disrupt the normal flow of web browsing.
The anticipated incentives for clients to use \sys are that it is both more automated and more privacy-preserving than current identity verification methods.

\section{Discussion} \label{sec:discussion}

\subsection{Multi-Device Support} \label{subsec:multiple_device_support}
Given the proliferation of smartphones, computers, and various IoT gadgets into many spheres of everyday life, 
we expect that most clients own (or soon will own) multiple devices with varying capabilities.
\sys{} has been designed with this scenario in mind. 
First, computational requirements of \sys{} can be met by any device that can establish TLS connections.
Any device that can perform a TLS handshake is sufficiently powerful to verify the signature on a \wrapper{}.
Second, storage is not an issue because \wrapper{s} can be stored anywhere. %
This also allows devices without a display to request \wrapper{s} and commit these to synced storage. 
Thereafter, any other device with an appropriate display owned by the same client can fetch these and perform data operations.
Third, the use of \texttt{BIP32} allows \sys to generate an arbitrary number of device public keys, which can be used to derive any number of \vcr{} public keys for interaction with different websites.
Since private keys are not stored on such devices, there is no threat against security of \sys, even if the number of devices increases.
Fourth, \vcr{s} do not need to be issued from the same device that originally interacted with the server.
Devices may need to be unprovisioned, but the client should still be able to issue \vcr{s} for interactions they made on those devices.
In \sys{}, the client can always transfer cookies and \wrapper{s} between devices.

\subsection{Multi-\vcr Support}
Unlinkability of \vcr{s} is critical for protecting consumers' privacy (as defined in Section~\ref{sec:system_and_threat_models}). 
However, it requires clients to generate and sign \vcr{s} for each session. 
To reduce overhead, a client can amend its key derivation mechanism.
For example, if a client prefers to use only one \vcr{} to refer to combined collected data for all sessions with a particular website, 
it can use the derivation path \texttt{t/i/s/j} where \texttt{t/i} is the derivation path of the device key as before, \texttt{s} is a server 
id and \texttt{j} is a server-specific (rather than global) session counter.
The client then collects all \wrapper{s} and generates a unified \vcr{} by signing it with the private key 
corresponding to the server \vcr{} public key (\texttt{t/i/s}). 
This server \vcr{} public key is then sent to the server.
The server derives \vcr{} public keys for individual sessions and verifies all \wrapper{s}.
For a new session with the same server, the client simply updates the server id (\texttt{s}) and repeats the process.

\subsection{Multi-Communication Protocol Support} \label{subsec:multi_comm_protocol}
So far, we focused on \sys{} being used over \texttt{HTTP}(\texttt{S}), the most common way to access Web services.
However, it can support any stateless protocol that assigns a unique identifier by using that identifier when generating a \wrapper{}.
Thus, as described in Section~\ref{sec:design:multi-application}, \sys is also applicable to applications that 
use other protocols to interact with online servers. Such applications can use \sys{} \wrapper{s} to bind client-generated 
public keys to any type of symmetric session identifier, and the same protocol to issue \vcr{s} for data associated with that identifier.

\subsection{Shared Devices} \label{subsec:shared_devices}
Certain types of client devices may be shared by multiple individuals, e.g., a smart TV used by all household members.
In the worst case, it may not be possible to associate usage data with a specific individual, e.g., if two or more people 
are using the device concurrently. This a general \emph{policy} question in the field of data protection; it is not unique to \sys{}.
There is no clear guidance in either GDPR or CCPA as to how to handle such situations.
However, \sys{} provides some mechanisms that could be used to assist with \emph{enforcing}, rather than \emph{defining}, 
data ownership policies for shared devices. For example, one conceivable data ownership policy for a shared device 
is that \emph{all} users of the device must consent to \vcr{s} being issued for sessions originating from this device.
In the example above, this would require all smart TV users to consent to a data access request, which may also 
help address the policy question of who actually \emph{owns} the data.
This could be achieved by using a signature scheme which requires all users participating in the creation of a \vcr{}.
The actual policy and procedures regarding data from shared devices should be defined 
by the regulators, while tools such as \sys{} should enable, rather than dictate, policy.

\subsection[3rd Party Storage]{$3^{rd}$ Party Storage} \label{subsec:third_party_storage}
One distinctive feature of \sys{}, as opposed to simply using cookies, is that possession of a \wrapper{} 
alone is insufficient to issue a \vcr{}. Thus, \wrapper{s} can be stored by third parties and retrieved only when needed.
This relaxes client-side storage requirements and creates a possible new business opportunity for (paid) service 
providers that manage \wrapper{s} on behalf of clients.

\subsection{Broad Identifier Support} \label{subsec:identifier_support}
\sys{} \wrapper{s} are a general means of binding client identification cookies to client-generated public keys.
Importantly, this method is \emph{non-invasive} and does not impose any constraints on the cookie, which is useful if a server changes its identification method, e.g., cookie content. \sys{} can also support future identification methods, as a server can simply issue \wrapper{s} that bind public keys to any new type of identifier, instead of the cookies.

\subsection[3rd-party Cookie Support]{$3^{rd}$-party Cookie Support} \label{subsec:third_party_cookie}
Third-party cookies are claimed to provide better, more personalized advertisements.
Although such cookies are commonly considered  detrimental to privacy~\cite{mayer2012third}, they are widely used; 
around $79$\% of $109$ million web pages include third-party cookies~\cite{median_cookie_size}.
\sys{} can support such cookies by modifying the browser extension to capture traffic going to third parties and extract and store all third-party cookies.
To obtain the \wrapper{s}, the client can either: (1) visit the third-party \wrapper{} endpoints individually, or (2) send all cookies to the first-party server, which would obtain \wrapper{s} on the client's behalf.
Note that \sys does not require enabling third-party cookies in order to function.
This is related to \emph{cookie syncing}~\cite{acar2014web,castelluccia2014,cookieMatching,urban2019study} in which, instead of placing multiple cookies on the client device, a set of servers associate the data they collect under a unified identifier.
\sys can likewise support this type of cookie and obtain the \wrapper{} by visiting the relevant endpoint.

\subsection{Further privacy considerations}
\label{subsec:further-privacy}
When a consumer issues a \vcr for a particular session, a potential risk arises that this action could reduce 
consumer privacy by allowing the service provider to link multiple sessions to the same consumer.
For example, there is a one-to-one mapping of a \vcr public key and a session, thus revealing \vcr keys 
to the servers might allow servers to link \vcr{s} as well.
To prevent this, we consider two approaches:

First, for data access requests, cryptographic techniques, such as Private Information Retrieval (PIR)~\cite{chor1995private}, 
can hide the identity (i.e., \vcr public key) of the data requested. For modify and delete operations, the problem is more challenging, 
because, if data is in plaintext, servers could detect what has been updated/deleted.
Furthermore, PIR is likely to incur a high bandwidth burden, since databases may retain data for very long periods of time (e.g., 10 years) and that large database might need to be sent to the client.

An exciting approach to overcome both these problems is to introduce Trusted Execution Environments (TEEs) on the server side. 
Using remote attestation~\cite{ias}, after ensuring that expected code is running on a server-side TEE, 
clients can create a secure channel to the TEE and send their \vcr keys over it.
The TEE can find the matching row in the database and return associated data.
This TEE-secured database can be populated with data collected during a secure connection between a client and a server, e.g., using techniques such as LibSEAL~\cite{Aublin2018}.
Server-side TEEs also allow servers to prove to clients how their data is used, also using remote attestation.
\vcr{} responses can be generated with such guarantees, providing more transparency and trust between clients and servers.

\subsection{Further Applications} \label{subsec:further_applications}
Although \sys{} focuses on \vcr{s} from accountless clients, it can also supplement verification of \vcr{s} from account-holding clients.
This can be useful, considering that passwords suffer from dictionary attacks and are often re-used on multiple servers.

\sys{} can also be used as a basis in scenarios that require client re-authentication.
For example, in the context of monetary transactions, receipts are currently used to prove that a client bought something from a 
merchant (server) in order to accept returns or perform exchanges.
With \sys{}, a client can supply a fresh \vcr{} public key during the purchase transaction and later generate a proof of 
ownership of the corresponding private key, which would anonymously confirm to the server that this is indeed the same customer.

\section{Related Work} \label{sec:related_work}

\taggedpara{Supporting VCR requests from accountless consumers}
Until now, the only means of authenticating accountless consumers have been {\em ad hoc}. \cite{pavur2019gdparrrrr} reports that 
such means may require one or more of: device cookies, government-issued IDs, signed and witnessed statements, utility bills, credit card numbers, 
or participation in a phone interview. However, these mechanisms are burdensome for consumers. Furthermore, they are insecure
(as shown in~\cite{pavur2019gdparrrrr,cagnazzo2019gdpirated,di2019personal,boniface2019security}), error prone (due to the manual processing) 
and privacy-invasive due to the additional information collected.
In contrast, \sys allows consumers to submit \vcr{s} in a secure and private manner without requiring any human interaction on the server side.

\taggedpara{Security of GDPR Subject Access Requests} As described in Section~\ref{sec:background}, the GDPR and CCPA grant subjects the rights to request access to their personal data collected by businesses, by submitting a \vcr{} or Subject Access Request (SAR).
Unfortunately, insecure (or easily circumventable) SAR verification practices open the door to potential leakage of personal data to unauthorized third parties.
Prior work~\cite{cagnazzo2019gdpirated, di2019personal, pavur2019gdparrrrr} has investigated various social engineering techniques for bypassing existing SAR verification practices.

Cagnazzo et al.~\cite{cagnazzo2019gdpirated} demonstrated that an unauthorized adversary can abuse the functionality provided by a business to update a victim subject's email and residential addresses. 
The adversary could then request access to \emph{``their''} data from this new address. Out of 14 organizations tested, 10 gave out personal information and 7 of these contained sensitive data.

Di Martino et al.~\cite{di2019personal} investigated the use of address spoofing techniques (e.g., using homoglyphs), 
as well as more sophisticated techniques such as manipulation of identity card images.
It was found that 15 out of 41 organizations with manual verification processes leaked personal data.
The remaining 14 organizations required an account-based login, which was impervious to such attacks, but is not available for the accountless consumers we consider in this work.

Pavur and Knerr~\cite{pavur2019gdparrrrr} performed an extensive evaluation of 150 companies' practices for SAR verification.
Results indicated that email address-based and account login were the most common, followed by device cookies, government IDs, and signed statements. 
Some organizations also requested utility bills, phone interviews, or credit card numbers. 
To bypass SAR verification, \cite{pavur2019gdparrrrr} created and sent a vague SAR letter to organizations.
Out of 150, 24\% disclosed personally-identifying information.

Boniface et al.~\cite{boniface2019security} also analyzed SAR verification practices for popular websites and third-party trackers.
The findings were that, in addition to possibly being insecure, SAR verification could undermine the privacy of subjects in order to verify the request.

\taggedpara{General Studies on GDPR Subject Access Requests} 
Urban et al.~\cite{urban2019your} performed a two-sided study of both data subjects and data-collecting organizations, with a focus on online advertising. 
For data subjects, consumer surveys were used to evaluate the usability of data transparency tools offered by the organizations, and to learn more about consumers' perceptions of these tools. 
\cite{urban2019your} also conducted surveys and interviews of organizations to get their views on the privacy regulations and business practices for SARs. 
The results paint a picture of discrepancy between the consumer' perspectives and the collected data, which is also corroborated by Ausloos and Dewitte~\cite{ausloos2018shattering}.
Furthermore, consumers seemed to show little interest in seeing raw technical data.
Similarly, Urban et al.~\cite{urban2019study} investigated SAR practices for online advertising companies and used cookie IDs to request collected data. 
This approach is similar to the symmetric approach in Section~\ref{sec:design_and_challenges}. 
\cite{urban2019study} reported that some companies requested ID cards or affidavits, while others directly used the cookie IDs in the browser.
Neither approach proves that the requestor is really the consumer about whom the data was collected.

Kr{\"o}ger~\cite{kroger2020appid} studied mobile applications and observed an even more fragile ecosystem with discontinued apps and disappearing consumer accounts while processing SARs. 
Another conclusion of this study was to move away from email-initiated and manual processes which are prone to errors.
In terms of compliance, the analysis by Herrmann and Lindemann~\cite{herrmann2016obtaining} showed 43\% compliance with access requests vs.\ 57\% compliance with deletion requests.

In addition to the above, Dabrowski et al.~\cite{dabrowski2019measuring} investigated cookie usage and how it is affected by privacy regulations, reporting that 11\% of EU-related websites set cookies for US-based consumers, though not for EU-based consumers.
Furthermore, up to 46.7\% of websites that appear in both the 2016 and 2018 Alexa top 100,000 sites stopped using persistent cookies without consumer permission. In the standardization realm, Zimmeck and Alicki~\cite{zimmeck2020standardizing} focused on ``Do Not Sell'' requests, which inform the websites that they may 
not share the consumer's information with third parties. \cite{zimmeck2020standardizing} developed a browser extension (OptMeowt) which conveys 
``Do Not Sell'' requests to the websites through headers and cookies.

\taggedpara{Asymmetric access tokens}
The most similar work to \sys{} from a technical perspective is Origin Bound Certificates (OBCs)~\cite{dietz2012origin} (also see RFC 8471~\cite{rfc8471}), which aims to strengthen TLS client authentication by converting cookies to asymmetric access tokens.
In OBC, the client generates a unique self-signed TLS client certificate for each website, in order to remain unlinkable across websites.
Although this does not authenticate the client to the website (due to the self-signed certificate), it does allow the server to ascertain whether this is the \emph{same} client from a previous interaction.
One benefit of this is that cookies can be bound to an OBC, such that, even if stolen, they cannot be used by an adversary.
This is very similar to how we bind \wrapper{s} to a client-generated key.
One key difference is that, in OBC, the cookies themselves are modified, whereas our use of \wrapper{s} means that \sys{} can be incrementally 
deployed on top of existing systems without needing to modify how they use cookies.
Another difference is that per-site certificates would not be suitable for accountless consumers, as these would allow servers to link together different visits from the same client. 
The alternative of generating per-TLS-session certificates introduces the key explosion problem, which we address in Section~\ref{sec:design_and_challenges}.
Finally, OBC and \sys{} differ in terms of their primary objectives: OBC aims to strengthen TLS channels in general, but is thus tightly coupled to the TLS protocol, whereas \sys{} aims to provide a specific mechanism for supporting \vcr{s}, which can be run over any communication protocol.

Another technology related to asymmetric access tokens is the FIDO Universal Authentication 
Framework (UAF)~\cite{fido-uaf}. UAF allows users to authenticate to servers using mechanisms other than passwords, e.g., biometrics.
It also supports multi-factor authentication, e.g., requiring both a PIN and a biometric.
The devices used to obtain such factors are called \emph{authenticators}.
During registration, authenticators generate and register a server-specific \emph{authentication key}, which they then use for subsequent authentications.
Whilst similar to the approach used in \sys, there are several reasons why the FIDO UAF protocol is not directly suitable for \sys.
First, \sys requires a fresh key pair to be generated for every \emph{session} per website, since these keys will be paired with session-specific cookies.
FIDO UAF does not meet this requirement, since only one key pair is generated per website.
Even if the FIDO UAF protocol were modified to generate a key pair per session per website, this would lead to the key explosion problem described in Section~\ref{subsubsec:key_generation}, and it would be particularly challenging to store all these keys in a resource-constrained device.
In contrast, in \sys, the use of BIP32 is critical to handle the significantly larger volume of keys and facilitate implementation on resource-constrained devices.
Second, using FIDO UAF would require trusted devices to be online when generating cookie wrappers, since both the public and private keys are generated by the trusted device.
This is not ideal for security conscious users who might prefer to keep their trusted device offline most of the time (e.g., in a locked safe).
In contrast, \sys supports both casual and security-conscious use cases by not requiring the trusted device to be present when generating public keys for cookie wrappers.

\section{Conclusions \& Future Work} \label{sec:conclusion_and_future_work}
Motivated by recent GDPR and CCPA regulations granting (even) accountless consumers rights to
access to data gathered about their behavior by web servers, we construct and evaluate \sys, a 
framework for authenticating accountless consumers. \sys is secure with respect to malicious clients and honest-but-curious servers,
easy to deploy, and imposes fairly low overhead.
Natural directions for future work include: 
(1) integration with client-side trusted execution environments (TEEs),
(2) more extensive support for the MODIFY \vcr type, and 
(3) support for unilateral server deletion of accountless-consumer data, which can occur if a server decides to delete consumer data without an explicit request.

\section*{Acknowledgements}
We thank NDSS reviewers for their valuable comments which helped improve this paper.
UCI authors were supported in part by funding from NSF Awards SATC-1956393
and CICI-1840197. The second author was also supported in part by The Nakajima Foundation.
The fourth author was supported in part by a US-UK Fulbright Cyber Security Scholar Award.

{
\balance 
\footnotesize
\raggedright
\bibliographystyle{IEEEtranS}
\bibliography{IEEEabrv,gdpr}
}

\normalsize

\raggedbottom

\pagebreak

\onecolumn

\section*{Formal Protocol Specification}
\label{sec:appendix-tamarin}

Listing~\ref{lst:tamarin} presents the formal model of the \sys protocol and the security properties verified using the Tamarin prover~\cite{meier2013tamarin}.
For details of the Tamarin syntax and conventions, please refer to the Tamarin prover website~\cite{tamarin-website}.  

\bigskip

\begin{lstlisting}[language=C,frame=single,caption={Tamarin specification of the messages exchanged in \sys{}, and the corresponding security lemmas.},label=lst:tamarin]
theory VICEROY
begin

builtins: signing, hashing

// Public key infrastructure
rule Register_pk:
  [ Fr(~skA) ]
  --[Unique($A)]->
  [ !Sk($A, ~skA), !Pk($A, pk(~skA)), Out(pk(~skA)) ]

rule Reveal_sk:
  [ !Sk(A, skA) ] --[ RevSk(A) ]-> [ Out(skA) ]


// Derivable asymmetric keys
rule Register_derived_key:
  [ !Sk($A, skA), Fr(~dskA) ]
  -->
  [ !Dsk($A, ~dskA), !Dpk($A, pk(~dskA)), Out(pk(~dskA)) ]

rule Reveal_dsk:
  [ !Dsk(A, dskA) ] --[ RevDsk(A) ]-> [ Out(dskA) ]


/* We formalize the following protocol

Trusted device (T)  |  Client device (C)  |  Web server (S)

[Optional] Trusted device generates master private key sk(t) and sends device public key pk(t/i) to the client's device [not included in the model since it is not mandatory to use a separate trusted device].

When the client begins interacting with a website, the website issues the client with a cookie.
0. S -> C: cookie

The client derives a fresh VCR public key pk(t/i/j) from the device public key and sends it to the server along with the cookie.
1. C -> S: cookie, pk(t/i/j) 

The server returns a cookie wrapper, which is a signature over the cookie and the provided VCR public key.
2. S -> C: sign_{sk(S)}{h(cookie, pk(t/i/j))}

When issuing a VCR, the user signs the request and cookie using the corresponding private key on their trusted device [not included in this model], and then sends the cookie, request, public key, wrapper, and signature to the server.
3. C -> S: cookie, request, pk(t/i), sign_{sk(S)}{h(cookie, pk(t/i/j))}, sign_{sk(t/i/j)}{h(request, cookie)}

If all the signatures can be verified, the server accepts the VCR and performs the requested operation. */

rule S_0:
    [ Fr(~cookie) ]
  --[  ]->
    [ Out( ~cookie ), !State_S_0($S, ~cookie) ]

rule C_1:
  let dpkC = pk(~dskC)
      m1 = <cookie, dpkC>
  in
    [ In(cookie), Fr(~dskC) ]
  --[ Request_wrapper( dpkC ) ]->
    [ Out( m1 ), State_C_1(~dskC, dpkC, cookie) ]

rule S_1:
  let m1 = <cookie, dpkC>
      m2 = sign(h(<cookie, dpkC>), skS)
  in
    [ !State_S_0($S, cookie), !Pk($S, pkS), !Sk($S, skS), In( m1 ) ]
  --[ Issue_wrapper($S, dpkC, <cookie>) ]->
    [ Out( m2 ), !State_S_1($S, pkS) ]

rule C_2:
  let request = <'op', ~nonce, pkS>
      m3 = <cookie, request, dpkC, wrapper, sign( h(<request, cookie>), dskC ) >
  in
    [ State_C_1(dskC, dpkC, cookie), !Pk($S, pkS), In(wrapper), Fr(~nonce) ]
  --[ Eq( verify(wrapper, h(<cookie, dpkC>), pkS), true )
    , Issue_VCR(dskC, $S, <cookie, request>) ]->
    [ Out( m3 ) ]

rule S_2:
  let request = <'op', nonce, pkS>
      m3 = <cookie, request, dpkC, wrapper, client_sig>
  in
    [ !State_S_1($S, pkS), In( m3 ) ]
  --[ Eq( verify(wrapper, h(<cookie, dpkC>), pkS), true )
    , Eq( verify(client_sig, h(<request, cookie>), dpkC), true )
    , Unique(nonce)
    , Accept_VCR($S, dpkC, <cookie, request>) ]->
    [  ]

restriction Equality:
  "All x y #i. Eq(x,y) @i ==> x = y"

restriction Uniqueness:
   "All x #i #j. Unique(x) @ i & Unique(x) @ j ==> #i = #j"

/* Wrapper unforgeability: whenever the server accepts a VCR from a client, then that server had previously issued a cookie wrapper to that client for the same cookie, or the adversary performed a long-term key reveal on the server, or the adversary knows the client's derived private key. */
lemma wrapper_unforgeability:
  " All server dskC cookie request #i.
        Accept_VCR(server, pk(dskC), <cookie, request>) @ i
      ==>
          (Ex #j. Issue_wrapper(server, pk(dskC), <cookie>) @ j & j < i)
        | (Ex #r. RevSk(server) @ r) | (Ex #r. KU(dskC) @ r) "

/* VCR unforgeability: whenever the server accepts a VCR, then the client with the corresponding private key issued that VCR, or the adversary performed a long-term key reveal on the server, or the adversary knows the client's derived private key. */
lemma VCR_unforgeability:
  " All server dskC cookie request #i.
        Accept_VCR(server, pk(dskC), <cookie, request>) @ i
      ==>
          (Ex #j. Issue_VCR(dskC, server, <cookie, request>) @ j & j < i)
        | (Ex #r. RevSk(server) @ r) | (Ex #r. KU(dskC) @ r) "

/* Replay resistance: the server will not accept a VCR for the same cookie and request combination more than once, unless the adversary knows the client's derived private key. */
lemma replay_resistance:
  " All server dskC cookie request #i #j.
        Accept_VCR(server, pk(dskC), <cookie, request>) @ i &
        Accept_VCR(server, pk(dskC), <cookie, request>) @ j
      ==>
          #i = #j | (Ex #r. KU(dskC) @ r) "

/* Consistency check: the server can accept a VCR without the adversary having performed a long-term key reveal on the server or knowing the client's derived private key. */
lemma accept_vcr_possible:
  exists-trace
  " Ex server dskC params #i.
      Accept_VCR(server, pk(dskC), params) @ i
    & not (Ex #r. RevSk(server) @ r)
    & not (Ex #r. KU(dskC) @ r) "

end
\end{lstlisting}

\end{document}